\documentclass{article}
\usepackage{spconf,amsmath,graphicx}

\usepackage{enumitem}
\setlist{nosep, leftmargin=14pt}

\usepackage{mwe} % to get dummy images

\usepackage{enumitem}

\usepackage{mwe} % to get dummy images

\usepackage{cite}
\usepackage{amsmath,amssymb,amsfonts}
\usepackage{algorithmic}
\usepackage{graphicx}
\usepackage{textcomp}
\usepackage{adjustbox}
\usepackage{comment}
\usepackage{multicol}
\usepackage{multirow}
\usepackage{xcolor}
\def\BibTeX{{\rm B\kern-.05em{\sc i\kern-.025em b}\kern-.08em
    T\kern-.1667em\lower.7ex\hbox{E}\kern-.125emX}}
\begin{document}

\title{Uncertainty Driven Bottleneck Attention U-net for Organ at Risk Segmentation
%\thanks{First two authors contributed equally}
}
\name{Abdullah Nazib$^1$, Riad Hassan$^2$, Zahidul Islam$^2$,Clinton Fookes$^1$}
\address{$^1$ SAIVT, Queensland University of Technology, Brisbane, Australia\\
	$^2$ Green University of Bangladesh\\
	{\tt\small nazib@qut.edu.au}}
%\name{Author Name$^{\star \dagger}$ \qquad Author Name$^{\star}$ \qquad Author Name$^{\dagger}$}
%
% \address{$^{\star}$ Affiliation Number One \\
%     $^{\dagger}$}Affiliation Number Two

\maketitle

\begin{abstract}
Organ at risk (OAR) segmentation in computed tomography (CT) imagery is a difficult task for automated segmentation methods and can be crucial for downstream radiation treatment planning. U-net has become a de-facto standard for medical image segmentation and is frequently used as a common baseline in medical image segmentation tasks. In this paper, we propose a multiple decoder U-net architecture and use the segmentation disagreement between the decoders as attention to the bottleneck of the network for segmentation refinement. While feature correlation is considered as attention in most cases, in our case it is the uncertainty from the network used as attention.      
For accurate segmentation, we also proposed a CT intensity integrated regularization loss. Proposed regularisation helps model understand the intensity distribution of low contrast tissues.  
We tested our model on two publicly available OAR challenge datasets. We also conducted the ablation on each datasets with the proposed attention module and regularization loss. Experimental results demonstrate a clear accuracy improvement on both datasets.
\end{abstract}

\begin{keywords}
Thoracic CT, Organs-at-Risk, Uncertainty, U-net.
\end{keywords}

\section{Introduction}
Radiation treatment planing requires meticulous delineation of organs that are adjacent to affected tissues. Delineating organs at risk (OAR) is tedious, time consuming and prone to inter-rater variability leading to unnecessary and dangerous radiation to OAR.
Recent advancements in deep learning (DL) have resulted many OAR segmentation methods. Automated OAR segmentation is a challenging task due to a number of factors that include patient to patient organ shape variations, low soft-tissue contrast, data or label imbalance etc. DL based models are highly accurate when organs are relatively large, consistent in shape and tissue contrast is high. 
Segmentation accuracy significantly falls for small organs with highly variable shape. 
Inter-rater variability in medical imaging is a common issue that introduces uncertainty into the training data, affecting model accuracy. Commonly used loss functions like Dice and Cross-Entropy (CE) focus on object shape and relative entropy, but they do not account for tissue contrast, which is crucial for precise boundary segmentation, especially in scenarios with low tissue contrast. 

To address aforementioned issues, we propose a multi-decoder U-net architecture where, like inter-rater variability, the segmentation disagreement between two decoders is considered as uncertainty and used to derive attention vector. The derived attention is used to refine bottleneck features for accurate segmentation. Considering low tissue contrast, a CT intensity integrated regularization loss is also proposed that considers tissue contrast in loss calculation. The proposed approach is evaluated on two public OAR segmentation datasets and compared with popular CNN based state-of-the-art segmentation methods.

\section{Related Works}
U-net become a standard model for medical image segmentation\cite{Ronneberger2015}. Many different variants of the U-net model have been developed in recent years \cite{Yagi2019,Kakeya2018,HEINRICH20191,Milletari2016,Wang2019,Wang2019b}. Most of these works have concentrated their effort in adding more layers or aggregating robust context information while processing the image in the layers of the U-net. In \cite{Ozan2018}, an attention gate is proposed for U-net, where the attention distribution is generated by processing the features from the deeper layer and shallower layers. \cite{Sinha2019MultiScaleSA} proposed guided multi resolution attention where attention is generated by aggregating multi-resolution features and is guided by a deep supervision loss. Similar deep feature based works also include \cite{Gu2021},\cite{YangLi2021} and \cite{Zhang2019} where authors proposed different attention mechanisms. 
We use three popular U-Net variants with attention, namely Attention U-Net, Unet++ and R2U-Net are selected as baseline to compare. The Attention U-Net \cite{Ozan2018} uses attention mechanisms in skip connections and decoder features to improve accuracy. The U-Net++\cite{zhou2020unet} create attention by concatenating dense, multi-scale skip features with decoder features. Lastly, the R2U-Net\cite{Alom2018} incorporates recurrent connections between encoding and decoding pathways to capture long-range dependencies to create attention while training.

\begin{figure*}[!htb]
\includegraphics[width=\linewidth,trim={0 88mm 0 20mm},clip]{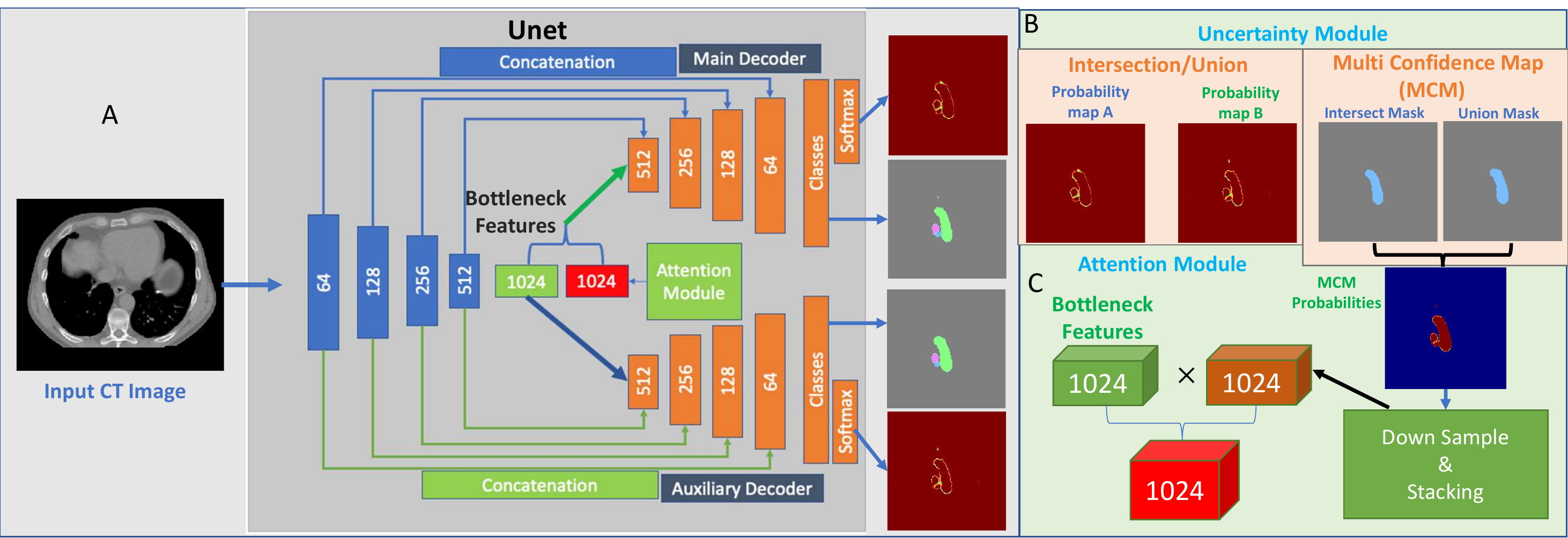}
\caption{Overview of the proposed approach.A) The U-net with auxiliary decoder. B) The Uncertainty module, C) Attention module.} \label{fig:Main}
\end{figure*}

\section{Method}
The proposed method is depicted in Figure \ref{fig:Main}. The approach utilizes a U-net architecture with two decoders,i) main decoder and ii)auxiliary decoder respectively. In auxiliary decoder, a perturbation layer is introduced, similar to the approach presented in \cite{ouali2020}. The perturbation layer employs random noise with a uniform distribution of $U(-0.3, 0.3)$ to modify the auxiliary decoder's output. Motivated by \cite{chen2021} \cite{Yang2022}, our the uncertainty module applies union and intersection operations producing union and intersection mask. The union mask takes all regions where both decoders agree and disagree, while the intersection mask takes only the regions where they agree. 
Similar to \cite{chen2021} and \cite{Yang2022}, a binary multi confidence map (MCM) is generated by taking all the regions of union and intersection masks (see Figure \ref{fig:Main}(B)). To extract the probabilities from the regions in the MCM, the maximum of the softmax function is applied to the output of each decoder, and the resulting values are multiplied with the multi-confidence mask. The equations for this procedure are as follows:
\begin{equation}
p_{main} = max(softmax(D_{main},dim=1))
\end{equation}
\begin{equation}
p_{aux} = max(softmax(D_{aux},dim=1))
\end{equation}
\begin{equation}
Attention = max(p_{main},p_{aux}) * MCM
%Attention =Normalize(P_{main} + P_{aux})*MCM
\end{equation}
Here, $D_{main}$ and $D_{aux}$ are the segmentation outputs from main and auxiliary decoders. $p_{main}$ and $p_{aux}$ are the corresponding softmax probabilities.  
The calculated attention is a single channel spatial attention where the probabilities of each pixel is obtained from the region of interest. To apply this attention to the bottleneck features, the spatial dimensions are downsampled and replicated along the channel dimension to match with dimension of the bottleneck. Finally, the uncertainty driven bottleneck attention (UDBA) is multiplied with the bottleneck features to select features that are required for the segmentation   (see Figure\ref{fig:Main}(C)).

\section{CT Intensity Integrated Regularization}
%Dice and cross entropy losses are commonly employed in medical image segmentation. The dice loss evaluates the agreement between the predicted segmentation and the actual segmentation. Thus, it primarily takes into account the shape of the object of interest. In contrast, the cross-entropy loss relies heavily on the distribution of the classes in the dataset. If the classes are evenly distributed, the classification results will be better. Since OAR segmentation requires precise boundary segmentation, the use of mere shape or class distribution sometime classify surrounding the boundary pixels as false positive. 
The CT intensities from the ground-truth and predicted regions are extracted using corresponding masks and absolute differences are measured. Thus, the calculated CTR helps the network learn tissue contrast more accurately. We also proposed a matrix version of CTR which compares both the intra-class and inter-class soft tissue contrast. 
The CT-intensity-integrated regularizer is
\begin{equation}
    \mathcal{L_{CTR}} = \frac{1}{WH}(|| \sum_{i=1}^{W}\sum_{j=1}^{H} CT*GT_{mask} - \sum_{i=1}^{W}\sum_{j=1}^{H} CT*P_{mask}||)
        \label{Eq:CTR}
\end{equation}
Let the number of classes are $\mathcal{N}$, the matrix version of CTR is a $\mathcal{N}\times \mathcal{N}$ matrix which is given by:
\begin{equation}
    \mathcal{M_{CTR}} = [\mathcal{CTR}_{i,j}]
    \label{Eq:CTRM}
\end{equation}
Each entry of the matrix is the CTR value between i-th class ground-truth and j-th class prediction, therefore inter-class CTR. When $i=j$ its the intra-class CTR. For simplicity we denote CTR matrix as CTRM. To optimize the network, the mean of the CTRM is taken and the resulting     
regularizer is given by
\begin{equation}
    \mathcal{L_{CTRM}} = \frac{1}{\mathcal{N}\mathcal{N}}\sum_{i=1}^{\mathcal{N}}\sum_{j=1}^{\mathcal{N}}[\mathcal{CTR}_{i,j}]
\end{equation}
Similar to \cite{ouali2020} we train the network with
three losses, two for segmentation and the one version of CTR in Eq.\ref{Eq:CTR} and \ref{Eq:CTRM}. Therefore, the training loss is
\begin{equation}
    \mathcal{L} = \mathcal{L}_{main} + \mathcal{L}_{aux} + \mathcal{L_{CTR}} or \mathcal{L_{CTRM}}
\end{equation}

\section{Experiments and Results}
\subsection{Datasets and Implementation}
The proposed regularization losses and attention module are evaluated on two datasets: SegThor and LCTSC. SegThor, with 40 CT training images, is collected from Hodgkin's lymphoma patients and has variable resolutions ($512\times512\times(150-284)$). Unfortunately, only a training set is available, so we split it into 35 for training and 5 for testing. LCTSC comprises 36 training and 24 testing CT images with organ annotations. In the absence of validation dataset, we apply 5 fold cross validation to prevent overfitting. Both datasets undergo the same preprocessing, involving the extraction of 2D axial slices from 3D volumes and contrast enhancement. The image slices are resized to $256\times256$. Three evaluation metrics, including Dice-Coefficient, Average Surface Distance (ASD), and Intersection Over Union (IoU), are used for fair quantitative assessment. The model is implemented using PyTorch (version 1.12.1) as a 2D model, focusing on the axial view during training. High-performance computing resources are utilized for training, including a 64GB RAM, A100 GPU, and a single-core 2.66GHz 64-bit Intel Xeon processor. The network is trained with the ADAM optimizer, a learning rate of 0.01, and a batch size of 1 for 200 training epochs.

%The evaluation of the proposed model was primarily based on the Dice coefficient and its standard deviation. Additionally, average surface distance (ASD) and intersection over union (IoU) are also employed for performance comparison between the proposed model and a baseline model. 

\subsection{Results on SegThor Dataset}
We compare the performance of our models with the attention U-net \cite{Ozan2018}, Unet++ \cite{zhou2020unet} and R2Unet \cite{Alom2018}. Table-\ref{Tab:SOTA_Segthor} presents the performance of our proposed method on the SegThor dataset. Notably, our CE+CTRM(UDBA) method excels in multiple metrics for various organs, with the best performance for the Heart. Unet++ and our Dice(UDBA) achieve high Dice scores for Esophagus and Aorta. AttUnet performs exceptionally well in ASD for the Esophagus but lags in other metrics. R2Unet generally under performs across all organs.overall findings suggest that proposed CE+CTRM(UDBA) is a promising method and achieves competitive results similar to the Unet++. 
\begin{table*}[ht]	
            \caption{\small Comparison between other baselines on SegThor Dataset. Note: CE=Cross-Entropy,
                UDBA=Uncertainty Driven Bottleneck Attention,
                CTR=CT Intensity Integrated Regularizer, CTRM= CTR Metrix }
		\label{Tab:SOTA_Segthor}
        \tiny
		\centering
    \begin{adjustbox}{width=\textwidth}
		\begin{tabular}{l |{c}| {c}| {c}| {c}|{c}|{c}| {c}| {c} |{c}| {c}| {c}| {c}}
			\hline
			\textbf{Method} &\multicolumn{3}{|c|}{\textbf{ Esophagus }} &\multicolumn{3}{c|}{\textbf{ Heart }} &\multicolumn{3}{c|}{\textbf{ Trachea }} & \multicolumn{3}{c}{\textbf{ Aorta}}\\
		    \cline{2-13}
             &\textbf{Dice}&\textbf{ASD}&\textbf{IoU} &\textbf{Dice}&\textbf{ASD}&\textbf{IoU} &\textbf{Dice}&\textbf{ASD}&\textbf{IoU} &\textbf{Dice}&\textbf{ASD}&\textbf{IoU}\\
             \hline
			Dice(UDBA)-Ours         &\textbf{0.81}&0.82&0.87   &0.92&1.11&0.78    &0.85&\textbf{0.46}&0.84   &0.92&0.73&0.91 \\
            \hline
            CE+CTRM(UDBA)-Ours      &0.74&1.45&0.89   &\textbf{0.95}&\textbf{0.94}&\textbf{0.93}  &\textbf{0.91}&0.57&\textbf{0.95}  &\textbf{0.93}&0.66&\textbf{0.93}\\
            \hline
            AttUnet\cite{Ozan2018}     &0.74&\textbf{0.66}&0.89 &0.80&1.57&0.61 &0.80&0.50&0.86 &0.88&0.59&0.88\\
            \hline
            Unet++\cite{zhou2020unet}   &\textbf{0.81}&0.82&\textbf{0.93}	&\textbf{0.95}&1.00&0.89	 &0.89&\textbf{0.46}&0.93	&\textbf{0.93}&\textbf{0.55}&0.91 \\
			\hline
            R2Unet\cite{Alom2018}  &0.69&0.86&0.83 	&0.77&1.97&0.63	 &0.81&0.50&0.83	 &0.81&0.90&0.83\\
            \hline
		\end{tabular}
  \end{adjustbox}
\end{table*}
\subsection{Results on LCTSC Dataset}
Similar to SegThor data, Table \ref{Tab:SOTA_LCTSC} presenting the comparison using Dice, ASD and IoU.
The proposed CE+CTRM(UDBA) shows strong performance overall, particularly excelling in terms of IoU scores for Esophagus, Spine, and Heart. Other proposed Dice+CTRM(UDBA) also demonstrates competitive results, especially for the Spine. AttUnet and Unet++ perform well across various organs, showcasing consistent performance. However, R2Unet generally performs lower compared to other methods. These findings suggest that proposed CE+CTRM(UDBA) and Dice+CTRM(UDBA) both are promising methods in comparison to Unet++ for organ segmentation on the LCTSC Dataset. Again, for LCTSC dataset, the challenge organizers kept the ledger board private. Hence we are unable to compare side-by-side with the other baseline methods.
\begin{table*}[!h]	
		\caption{\small\centering Comparison between other baselines on LCTSC Dataset.}
		\label{Tab:SOTA_LCTSC}
		\centering
        \begin{adjustbox}{width=\textwidth}
		\begin{tabular}{l |{c}| {c}| {c}| {c}|{c}|{c}| {c}| {c} |{c}| {c}| {c}| {c}| {c}| {c}| {c}}
			\hline
			\multirow{2}{*}{\textbf{Method}} &\multicolumn{3}{|c|}{\textbf{Esophagus}} &\multicolumn{3}{c|}{\textbf{Spine}} &\multicolumn{3}{c|}{\textbf{Heart}} &\multicolumn{3}{c}{\textbf{Lung(L)}} &\multicolumn{3}{|c}{\textbf{Lung(R)}}\\
		    \cline{2-16}
             &\textbf{Dice}&\textbf{ASD}&\textbf{IoU} &\textbf{Dice}&\textbf{ASD}&\textbf{IoU} &\textbf{Dice}&\textbf{ASD}&\textbf{IoU} &\textbf{Dice}&\textbf{ASD}&\textbf{IoU} &\textbf{Dice}&\textbf{ASD}&\textbf{IoU}\\
             \hline
			 CE+CTRM(UDBA)-ours      &0.63&1.76&\textbf{0.80} &0.88&0.71&\textbf{0.91}  &\textbf{0.92}&\textbf{1.35}&\textbf{0.84} &\textbf{0.97}&0.65&\textbf{0.92}  &\textbf{0.97}&0.71&\textbf{0.92}\\
            \hline
            Dice+CTRM(UDBA)-ours     &0.71&1.58&0.75	    &\textbf{0.89}&\textbf{0.66}&0.89	 &0.88&1.78&0.50	    &0.95&\textbf{0.62}&0.85	&0.96&\textbf{0.68}&0.89\\
            \hline

            AttUnet\cite{Ozan2018}     &0.66&\textbf{1.09}&0.70     &\textbf{0.89}&\textbf{0.66}&\textbf{0.91}   &0.69&3.26&0.39   &0.95&0.68&0.86   &0.96&0.72&0.90\\
            \hline
            Unet++\cite{zhou2020unet}  &\textbf{0.72}&1.14&0.71	&0.89&0.66&0.91	 &0.89&1.63&0.51	&0.96&0.69&0.87	&0.96&0.73&0.89\\  
			\hline
        
            R2Unet\cite{Alom2018} &0.51&1.74&0.54	&0.92&0.65&0.89	 &0.55&3.34&0.31	    &0.94&0.73&0.78	    &0.95&0.76&0.78\\
            \hline
		\end{tabular}
  \end{adjustbox}
\end{table*}

\subsection{Ablation Study}
To evaluate the effectiveness of proposed uncertainty driven attention module (UDBA) and CTR, twelve different experiments are conducted and presented on Table-\ref{Tab:segthor} and Table-\ref{Tab:LCTSC}. Each loss and loss combination is tested with and without UDBA. The accuracy is measured by mean Dice of each organ from the respective dataset.

Table-\ref{Tab:segthor} reports the dice accuracy of U-nets on the SegThor dataset for four organs - Esophagus, Heart, Trachea, and Aorta.
It is clear from the Table-\ref{Tab:segthor} data, that the different loss combination performs differently on four different organs. For esophagus, the model with dice and UDBA achieves the top score with 0.81$\pm$0.045. For heart, the CE and its combinations are leading  compared to the dice losses. The CE(UDBA) and CE+CTRM(UDBA) achieves the 0.95$\pm$0.008 and 0.95$\pm$0.007 for the heart. Same pattern goes for the trachea. In trachea, the CE+CTRM(UDBA) achieves the top dice score with 0.91$\pm$0.022. In aorta, the dice scores of dice loss variants fluctuates between 0.90 to 0.92 while the CE variants are fluctuating from 0.89 to 0.93. Again, in aorta the CE+CTRM(UDBA) scores the highest dice score with 0.93$\pm0.013$. From the Table-\ref{Tab:segthor} it is clear that the UDBA module systematically improves the accuracy regardless of loss combinations.
The LCTSC dataset has five different regions of interest, including Esophagus, Spinal Cord, Heart, left and right Lung. 

In Table-\ref{Tab:LCTSC}, Dice+CTRM(UDBA) achieves the top score for esophagus and spinal cord with 0.71$\pm$0.063 and 0.89$\pm$0.009, respectively. For heart and lungs, the CE(UDBA) achieves the highest scores in all three regions with the scores 0.92$\pm$0.033, 0.97$\pm$0.003 and 0.97$\pm$0.003. The combination
CE+CTRM (UDBA) also achieves very similar scores to CE(UDBA) in four organs except the esophagus.

\begin{figure}[!htb]
    \centering
        \includegraphics[width=\linewidth]{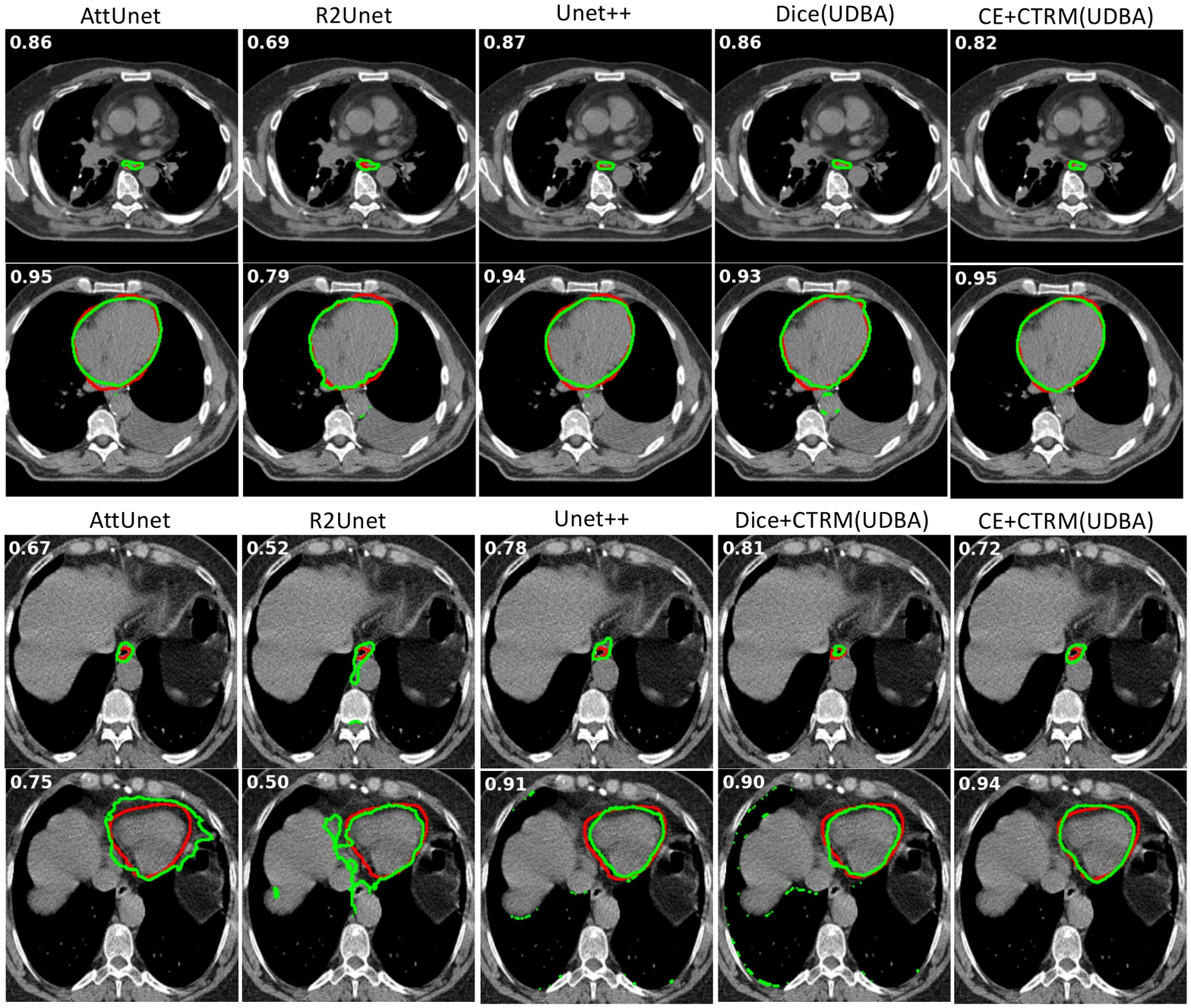}
    \caption{Sample segmentation images from different methods on SegThor (top two rows) and LCTSC dataset (bottom tow rows). The organs are Esophagus and Heart from both datasets. Red=Ground-truth, Green=Prediction.}
    \label{fig:segthor}    
    \end{figure}

In Table-\ref{Tab:LCTSC}, dice-based models excel in esophagus and spinal cord, while CE-based models perform better for three other organs. The UDBA module consistently enhances accuracy across loss combinations. 
Figure-\ref{fig:segthor} showcases sample outputs from different models on both datasets. The red and green contours representing ground-truth and model predictions. Only axial slices from 3D volumes were considered for simplicity.

\begin{table}[!htb]
    \caption{\small Dice Accuracy of Unets with and without attention module on Segthor Dataset. Combination of different losses and proposed regularizations are presented.}
    \label{Tab:segthor}
	\centering
    \begin{adjustbox}{width=\linewidth}
            \begin{tabular}{l |{c}| {c}| {c}| {c}}
			\hline
			\textbf{Losses (Unet)} &\textbf{ Esophagus } &\textbf{ Heart } &\textbf{ Trachea } &\textbf{ Aorta}\\
		    \hline
                \hline
			     Dice           &$0.74$            &$0.93$             &$0.85$            &$0.91$ \\
                Dice(UDBA)       &$\textbf{0.81}$   &$0.93$    &$0.84$   &$0.92$ \\
                Dice+CTR         &$0.72$   &$0.92$    &$0.84$   &$0.90$ \\
                Dice+CTR(UDBA)   &$0.76$   &$0.93$   &$0.87$    &$0.92$ \\
                Dice+CTRM        &$0.75$   &$0.93$   &$0.84$    &$0.90$ \\
                Dice+CTRM(UDBA)  &$0.74$   &$0.93$   &$0.87$    &$0.92$ \\
                CE               &$0.71$   &$0.94$   &$0.88$    &$0.89$ \\
                CE(UDBA)         &$0.78$   &$\textbf{0.95}$    &$0.90$  &$0.92$ \\
                CE+CTR		     &$0.70$   &$0.93$    &$0.89$	 &$0.88$\\
                CE+CTR(UDBA)	 &$0.74$   &$0.94$	  &$0.90$	 &$0.92$\\
                CE+CTRM		     &$0.69$	 &$0.93$  &$0.88$	   &$0.86$\\
                CE+CTRM(UBDA)	 &$0.74$	 &$\textbf{0.95}$	   &$\textbf{0.91}$	  &$\textbf{0.93}$\\
			\hline
		\end{tabular}
  \end{adjustbox}
\end{table}
\begin{table}[!htb]	
		\caption{\small Dice Accuracy of Unets with and without attention module on LCTSC  Dataset. Combination of different losses and proposed regularization are presented.}
		\label{Tab:LCTSC}
		\centering
        \begin{adjustbox}{width=\linewidth}
		\begin{tabular}{l |{c}| {c}| {c}| {c}| {c}}
			\hline
			\textbf{Losses (Unet)} &\textbf{Esophagus} &\textbf{Spine} &\textbf{Heart} &\textbf{Lung(Left)} &\textbf{Lung(Right)}\\
		    \hline
                \hline
			Dice            &$0.66$ &$0.88$ &$0.71$ &$0.94$ &$0.95$ \\
            Dice(UDBA)          &$0.68$ &$0.88$ &$0.82$ &$0.96$ &$0.95$ \\
            Dice+CTR            &$0.63$ &$0.88$ &$0.81$ &$0.93$ &$0.95$ \\
            Dice+CTR(UDBA)      &$0.68$ &$0.88$ &$0.47$ &$0.93$ &$0.94$ \\
            Dice+CTRM           &$0.65$ &$0.88$ &$0.78$  &$0.93$ &$0.94$ \\
            Dice+CTRM(UDBA)     &$\textbf{0.71}$ &$\textbf{0.89}$ &$0.88$ &$0.95$ &$0.96$\\
            CE                  &$0.61$ &$0.86$ &$0.89$ &$0.96$ &$0.96$ \\
            CE(UDBA)            &$0.63$ &$0.88$ &$\textbf{0.92}$ &$\textbf{0.97}$ &$\textbf{0.97}$ \\
            CE+CTR              &$0.56$ &$0.86$ &$0.90$ &$0.96$ &$0.96$\\
            CE+CTR(UDBA)        &$0.62$  &$0.87$ &$0.91$ &$0.96$ &$0.97$\\
            CE+CTRM             &$0.52$  &$0.86$ &$0.90$ &$0.96$ &$0.96$\\
            CE+CTRM(UDBA)       &$0.63$ &$0.88$  &$\textbf{0.92}$ &$0.96$ &$0.97$\\
			\hline
		\end{tabular}
   \end{adjustbox}
\end{table}

\section{Discussion and Conclusion}
In this work, we presented the utilization of slightly different network segmentation outputs to estimate uncertainty. We also demonstrated that estimated uncertainty can be used as attention to enhance segmentation accuracy. we have created a simple, easy-to-implement 2D U-net architecture that is trained exclusively with axial views. The application of uncertainty-based attention for improved segmentation was validated through experiments on two OAR segmentation datasets.

We also introduced the CT intensity integrated regularization loss to help the network learn texture distributions of organs and their shapes. Experimental evaluation confirmed efficiency of intensity integration for difficult-to-segment organs. Future research will focus on further improving the accuracy of proposed regularizer near the boundary of closely located organs.

\section{COMPLIANCE WITH ETHICAL STANDARDS}
This research study was conducted retrospectively using human subject data made available in open access through the
LCTSC and SegThor dataset. Ethical approval was not required.
\section{Declaration of Competing Interest}
The authors declare that they have no known competing financial interests or personal relationships that could have appeared to influence the work reported in this paper.
\section{Acknowledgment}
The work is supported by QUT High Performance
Computing (HPC) facility.
\bibliographystyle{elsarticle-num}
\bibliography{references}

\end{document}